\documentclass[superscriptaddress,showpacs,jmp,twocolumn,amsmath,amssymb]{revtex4-1}

\usepackage{graphicx}
\usepackage{dcolumn}
\usepackage{bm}
\usepackage{textcomp}
\usepackage{verbatim} 
\usepackage{amsmath}
\usepackage{eqnarray}

\usepackage[utf8]{inputenc}

\usepackage{datetime}

\usepackage{color}
\definecolor{darkblue}{rgb}{0,0,0.6}
\usepackage[linkcolor=darkblue, citecolor=darkblue,colorlinks=true, breaklinks=true]{hyperref}

\begin{document}

\preprint{AIP/123-QED}

\title{Quantification of Ultrafast Nonlinear Photothermal and Photoacoustic Effects in Molecular Thin Films via Time-Domain Brillouin Scattering}

\author{Valentin Cherruault}
\affiliation{Institut de Physique de Rennes, UMR CNRS 6251, Université de Rennes, 35042 Rennes, France}
\author{Franck Camerel}
\affiliation{Institut des Sciences Chimiques de Rennes, UMR CNRS 62226, Université de Rennes, 35042 Rennes, France}
\author{Julien Morin}
\affiliation{Institut de Physique de Rennes, UMR CNRS 6251, Université de Rennes, 35042 Rennes, France}
\author{Amédée Triadon}
\affiliation{Institut des Sciences Chimiques de Rennes, UMR CNRS 62226, Université de Rennes, 35042 Rennes, France}
\affiliation{Research School of Chemistry, Australian National University, Canberra, ACT 2601, Australia}
\author{Nicolas Godin}
\affiliation{Institut de Physique de Rennes, UMR CNRS 6251, Université de Rennes, 35042 Rennes, France}
\author{Ronan Lefort}
\affiliation{Institut de Physique de Rennes, UMR CNRS 6251, Université de Rennes, 35042 Rennes, France}
\author{Olivier Mongin}
\affiliation{Institut des Sciences Chimiques de Rennes, UMR CNRS 62226, Université de Rennes, 35042 Rennes, France}
\author{Jean-François Bergamini}
\affiliation{Institut des Sciences Chimiques de Rennes, UMR CNRS 62226, Université de Rennes, 35042 Rennes, France}
\author{Antoine Vacher}
\affiliation{Institut des Sciences Chimiques de Rennes, UMR CNRS 62226, Université de Rennes, 35042 Rennes, France}
\author{Mark G. Humphrey}
\affiliation{Research School of Chemistry, Australian National University, Canberra, ACT 2601, Australia}
\author{Maciej Lorenc}
\email{maciej.lorenc@cnrs.fr}
\affiliation{Institut de Physique de Rennes, UMR CNRS 6251, Université de Rennes, 35042 Rennes, France}
\author{Frederic Paul}
\email{frederic.paul@cnrs.fr}
\affiliation{Institut des Sciences Chimiques de Rennes, UMR CNRS 62226, Université de Rennes, 35042 Rennes, France}
\author{Thomas Pezeril}
\email{thomas.pezeril@cnrs.fr}
\affiliation{Institut de Physique de Rennes, UMR CNRS 6251, Université de Rennes, 35042 Rennes, France}

\date{\today}		

\begin{abstract}

Improving the efficiency of photothermal (PT) therapies and photoacoustic (PA) imaging at the microscopic scale hinges on developing multiphoton-absorbing photothermal molecules or contrast agents that operate in the near-infrared (NIR) range. These advanced agents or molecules will enable excitation with NIR lasers, in an improved transparency range for biological tissues, while enabling minute, highly localized spatial control of the excitation area. However, progress in this field requires innovative experimental techniques to characterize photothermal and photoacoustic effects under multiphoton excitation. In this article, we showcase a study of a model organometallic molecular compound excited via two-photon absorption (2PA) using femtosecond laser pulses. Based on a time-domain Brillouin scattering technique, well adapted for investigating ultrafast nonlinear optical absorption processes in ultrathin films on substrates, we determine the effective nonlinear absorption coefficients of the compound directly linked to PT/PA. Our findings provide a practical approach for exploring and optimizing nonlinear PT/PA absorbers and contrast agents.
 
\end{abstract}

\maketitle


\section*{Introduction}

Photoacoustic imaging has emerged as one of the fastest-growing fields in molecular bioimaging over the last decade \cite{Merkes2020}. This imaging technique relies on the unique capabilities of “contrast agents,” which are photothermal substances that absorb light, convert the captured photon energy into heat \cite{Hu2021, Xu2022}, and release acoustic waves through rapid thermal expansion. Upon photoexcitation, these acoustic waves propagate through the medium and can be detected at various locations. By applying sonar-inspired principles to analyze and process the detected signals, it becomes possible to reconstruct detailed 3D images of biological tissues surrounding the emitting sources \cite{Merkes2020, Fu2019}.  

Despite significant technical advances in image processing, photoacoustic microscopy still faces substantial challenges, particularly rapid resolution losses when imaging deeper tissues. Anticipating that the evolution of contrast agents might follow a similar trajectory to photosensitizers used in photodynamic therapy \cite{Dreano2023}, researchers have recently proposed transitioning from one-photon absorbing (1PA) contrast agents to two-photon absorbing (2PA) ones \cite{Hu2020}. The net result of such a shift is not only to shift the photoexcitation energy to longer wavelengths --typically within a spectral range where biological tissues are more transparent--thereby allowing the laser beam to penetrate deeper into samples, but also to minimize collateral photodamage. This is because multiphoton absorption is confined to the focal point of the excitation laser beam, unlike the broader absorption profile seen in 1PA processes \cite{Dreano2023}. Furthermore, the exquisite 3D spatial control and nonlinear intensity dependence of the 2PA process are expected to significantly enhance imaging contrast compared to the more conventional one-photon approach \cite{Hu2020, Urban2014, Nedosekin2014}. However, despite a few promising experiments in this domain \cite{Hu2020, Urban2014, Langer2013}, further work is needed to translate these expectations into practical outcomes. It is worth noting that many techniques currently employed to study linear PT \cite{Gu2013, Lucas2022, Wang2014, Roper2007} or PA effects \cite{Braslavsky1992} are often ill-suited for investigating their nonlinear counterparts due to the highly localized nature of multiphotonic excitation. 

In this context, the development of experimental methods capable of characterizing and optimizing nonlinear PT and PA phenomena is highly desirable to accelerate the design of advanced multiphotonic contrast agents. There remains an urgent need to identify high-performance 2PA chromophores for nonlinear photoacoustic imaging, as most clinically used contrast agents are still based on 1PA mechanisms. Unfortunately, techniques that enable accurate screening of 2PA-absorbing chromophores through direct monitoring of their nonlinear photothermal or photoacoustic effects are still scarce \cite{Terazima2002, Tam1979}.  

\begin{figure*}[!ht]
\centering
\includegraphics[width =\textwidth]{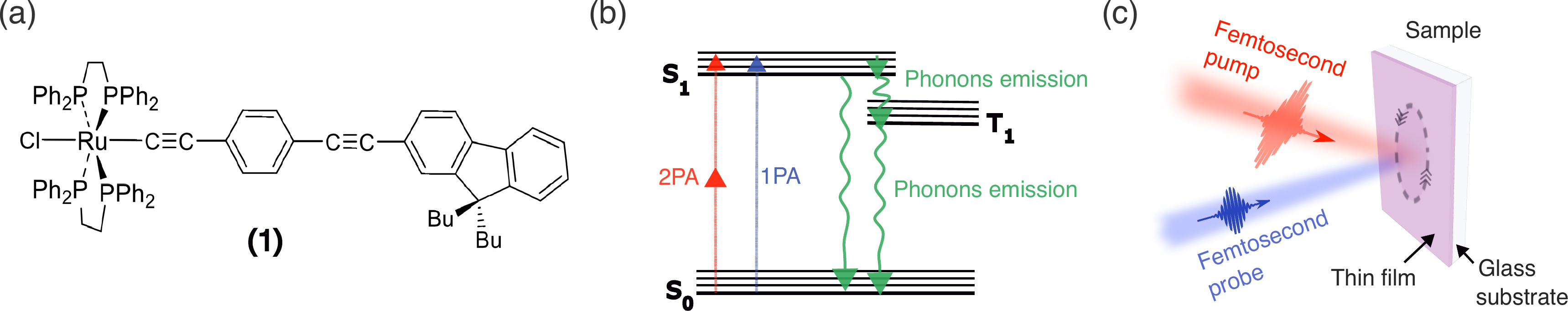}
\caption{(a) Organoruthenium compound \textbf{1} used in the study. (b) Classical deactivation pathways for a singlet excited state populated by one- (1PA) or two-photon (2PA) absorption and non-radiative relaxation that gives rise to heat release and phonons emission. (c) Sketch of the spinning femtosecond pump-probe setup depicting the circular laser trajectory on the sample surface.}
\label{fig:organoruthenium}
\end{figure*}

In this article, we present a modified femtosecond pump-probe technique tailored for investigating 2PA properties of molecules organized in ultrathin film layers on substrates. The ultrafast 2PA process is characterized through time-domain Brillouin scattering, providing insights into the laser absorption dynamics of the compound. The technique draws inspiration from our previous work \cite{Chaban2022}, where multi-photon absorption was detected in nanocrystals. However, unlike our study\cite{Chaban2022} of a quasi-3PA (sequence of 2PA and 1PA from the excited state), our results focus now on a model organometallic molecule \textbf{1} presenting a textbook 2PA absorption band at the NIR pump wavelength and with a fairly large 2PA cross-section\cite{Triadon2018}. Additionally, we outline a streamlined method to accurately extract at once the nonlinear optical parameters of interest for this compound by combining 1PA and 2PA measurements, eliminating the need for additional assessments like heat capacity or thermal expansion measurements, but providing instead key data on its PT and PA effects at very short timescales.

\section*{Methods}


The investigated complex \textbf{1}, spin-coated as an homogenous thin-film on a fused silica substrate, is the known organoruthenium molecular compound depicted in Fig.~\ref{fig:organoruthenium}(a). The film thickness at the vicinity of the laser-excited area is of 100~nm, with a thickness distribution or roughness below 5~nm and no apparent polycrystallinity (see Supplementary for more details). {\color{black}Its relatively large 2PA cross-section expressed in Goppert-Mayer (GM) units $\sigma_2$~=~1075~$\pm$~190~GM \cite{Triadon2018} at a 800~nm laser wavelength (with 1 GM equal to 10$^{-50}$cm$^4$s), makes it an ideal molecular “model” for testing the propensity of time-domain Brillouin scattering to study nonlinear PT and PA effects.} In the following, we demonstrate how to extract the effective nonlinear absorption coefficient $\alpha^{(2)}$ from time-domain Brillouin scattering. The coefficient is linked to the 2PA cross section $\sigma_2$ (expressed in units of Goppert-Mayer (GM)) through the relationship \cite{Liaros2017},
\begin{equation}
\alpha^{(2)}=\sigma_2 \, N_0 / E
\label{sigma}
\end{equation} where $N_0$ is the molecular density of the compound (number of molecules per unit volume) and $E$ the photon energy.

For diamagnetic, non-emissive molecules such as the one investigated in this study, heat release into the surrounding medium is driven by various non-radiative relaxation processes and the simultaneous generation of acoustic phonons. While the latter phenomenon has received limited attention since its initial exploration in the late 1960s, it is evident that not all non-radiative relaxation pathways contribute equally to the formation of acoustic waves, as their kinetics and temporal ordering differ\cite{Hu2021,Xu2022}. The overall thermal relaxation process is traditionally conceptualized as depicted in Fig.~\ref{fig:organoruthenium}(b) \cite{Xu2022}. Briefly, after photoexcitation of the molecule into its first absorption band, the resulting Franck-Condon state undergoes a fast vibrational cooling into the singlet state S$_1$ (S$_n$), followed by internal conversion into the ground state S$_0$. A competing pathway involves stepwise decay to S$_0$ via the first triplet state T$_1$, initiated by intersystem crossing from S$_n$ to the triplet manifold, followed by vibrational cooling to T$_1$. Subsequently, reverse intersystem crossing returns the molecule to vibrationally excited ("hot") S$_0$, followed by vibrational cooling. For many organic molecules, this instersystem crossing-mediated pathway is kinetically much slower and often irrelevant to the buildup of acoustic waves \cite{Terazima2002}, as it leads to temporally delayed heat release. In short, the mechanism of heat release from a photoexcited molecule depends on the relative kinetics of competing internal conversion and instersystem crossing processes. Additionally, vibrational cooling efficiency is determined by the availability and nature of vibrational modes that facilitate energy transfer to the surrounding medium, typically a solvent acting as the heat acceptor.  

\begin{figure*}[!htb]
    \centering
    \includegraphics[width =\textwidth]{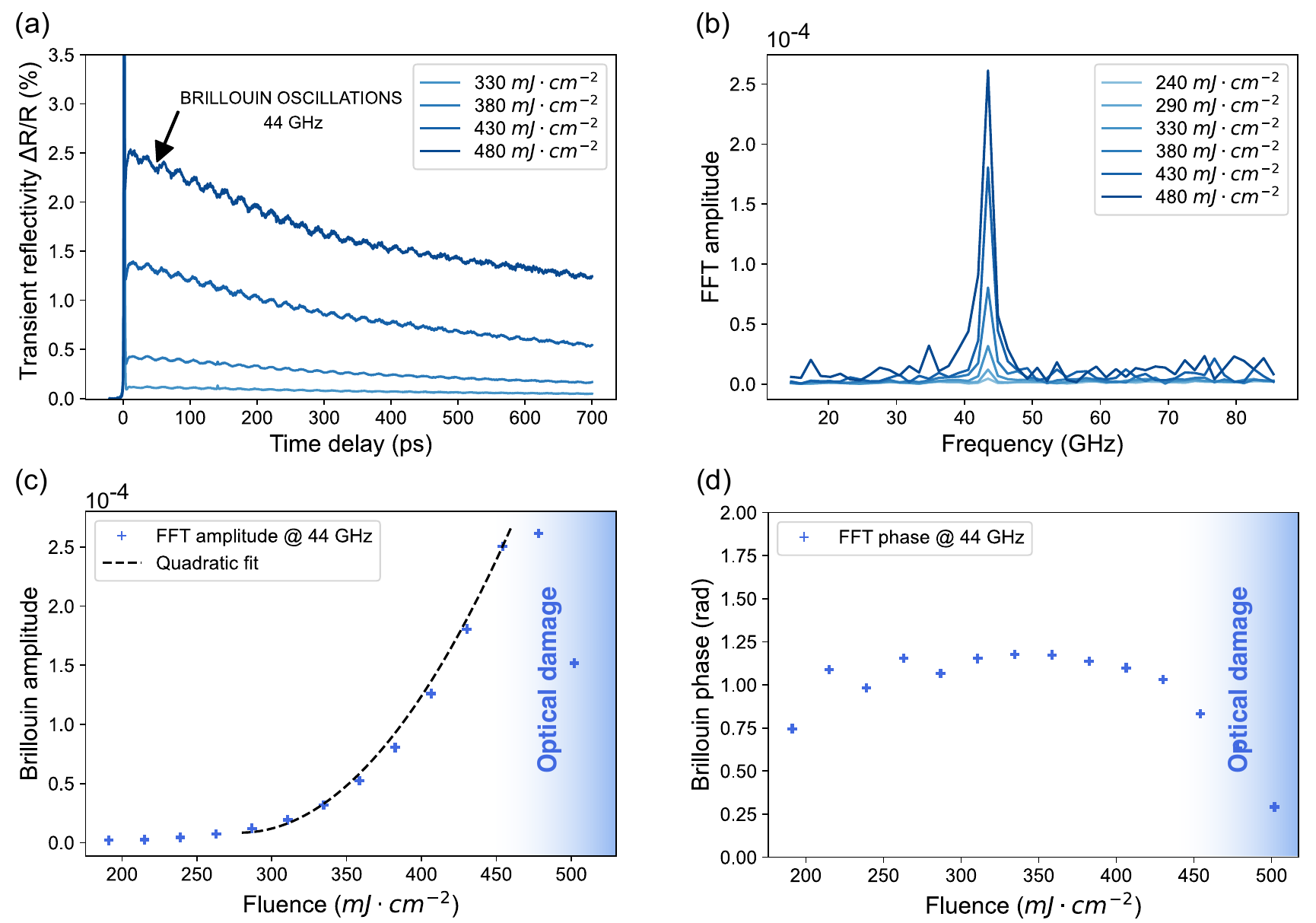}
    \caption{TDBS data at varying pump fluences at a pump wavelength of 780~nm. (a) Time-domain transient reflectivity data. (b) FFT of the the transient reflectivity data after slow background removal. (c) FFT amplitude of the Brillouin peak at 44~GHz versus pump fluence. (d) FFT phase of the Brillouin peak at 44~GHz versus pump fluence.}
    \label{fig:rouge_bleue_results}
\end{figure*}

In this study, time-domain Brillouin scattering (TDBS) is employed to investigate these non-radiative processes on femtosecond to picosecond timescales, providing insights into the energy dissipation mechanisms that contribute to PT and PA effects. Our TDBS experiment has been thoroughly detailed in earlier work \cite{Chaban2022}. In essence, the setup uses a femtosecond Ti-sapphire Coherent RegA 9000 regenerative amplifier, emitting 250~fs pulses at a 250~kHz repetition rate, centered at 780~nm with a maximum pulse energy of 4~$\mu$J. The beam is split into pump and probe beams, with the pump beam passing through a motorized delay stage for temporal control between the pump-probe beams on the sample. A key feature of the setup is the spinning wedge mirror, rotating at 50~Hz, which enables pump-probe measurements at alternating locations on the sample, mimicking single-shot pump-probe experiments. This method is crucial for preventing cumulative heating or laser damage during measurements obtained at high laser fluences as in this paper. Both the pump (14~$\mu$m at 1/e$^2$ for the 780 nm and 10~$\mu$m at 1/e$^2$ for the 390 nm) and probe (8~$\mu$m at 1/e$^2$) beams are spatially overlapped on the sample surface and scanned thanks to the spinning mirror in a circular trajectory of $\sim$200~$\mu$m in diameter, as sketched in Fig.~\ref{fig:organoruthenium}(c).

The organoruthenium molecular compound is totally transparent at the laser pump wavelength of 780~nm but exhibits a very strongly absorbing near-UV band corresponding to 2PA at half the pump wavelength of 390~nm, see Supplementary information. {\color{black}In fact, regardless of whether laser energy absorption in the thin film occurs via single-photon or multiphoton processes, it results in an ultrafast local temperature rise throughout the entire film thickness. This rapid thermal expansion generates an acoustic wavepacket, whose spatial profil closely matches the full film thickness, that travels through the film and is simultaneously partially transmitted into the glass substrate at each propagation cycle. As the acoustic wave propagates, it is detected by the time-delayed probe pulse using TDBS.} The overall probe beam, reflected from the sample interface and scattered from the acoustic wave packet, is directed to a photodiode in order to measure transient differential reflectivity $\Delta$R/R as a function of time delay between pump and probe pulses.

\section*{Results and discussions}

In order to monitor the nonlinear excitation of propagating strains caused by nonlinear optical absorption in the film, a series of TDBS measurements across a wide range of pump fluences were conducted. Fig.~\ref{fig:rouge_bleue_results}(a) presents the TDBS data recorded at varying pump fluences at a pump wavelength of 780~nm for which the available laser energy is the highest. The TDBS data exhibit several characteristic features. First, there is a pronounced spike near zero time delay, which corresponds to the absorption of the pump photons by the electrons in the material, leading to rapid changes in the compound optical properties and a strong reflectivity change of the probe light. This is followed by a long-lasting background signal, spanning hundreds of picoseconds, which represents the relaxation of the excited electrons back to the ground state (S$_0$) as illustrated in Fig.~\ref{fig:organoruthenium}(b). The sharp rise in the background close to time zero indicates a fast relaxation from the excited state to the singlet state (S$_1$), while the flatter, prolonged background likely corresponds to the slower relaxation from the first triplet (T$_1$) to the ground state, which in the case of \textbf{1} might kinetically dominate internal conversion from the S$_1$ state given the existence of triplet ligand field states at low energy\cite{Triadon2018}. Both of these relaxation processes result in the release of heat and the generation of acoustic phonons within the compound. Superimposed on this electronic transitions background are oscillations that arise from the propagation of the acoustic wavepacket into the transparent glass substrate, see Fig.~\ref{fig:rouge_bleue_results}(a). These oscillations, known as Brillouin frequency oscillations, are characterized by a frequency $\nu_B$, which is linked to the velocity of the acoustic waves $v$ and the refractive index $n$ of the glass substrate, following the well-known relation \cite{Brillouin_scattering}, 
$$\nu_B = \frac{2nv}{\lambda},$$
where $\lambda$ is the probe wavelength. The refractive index of the glass substrate at 390~nm probe wavelength is well known, as is its sound velocity, so the $\sim$44~GHz frequency peak highlighted in the fast Fourier transform (FFT) of Fig.~\ref{fig:rouge_bleue_results}(b), can be unambiguously identified as the Brillouin frequency of fused silica. 

Since the thin film is transparent at the pump wavelength, the Brillouin signals observed in the experiment must originate from nonlinear optical absorption processes, which exhibit a nonlinear dependence on the pump fluence. As shown in Fig.~\ref{fig:rouge_bleue_results}(c), the Brillouin FFT amplitude increases in a highly nonlinear manner with increasing laser pump fluence. Specifically, the amplitude demonstrates a quadratic dependence on the pump fluence, indicating a second-order response. This quadratic relationship becomes evident above a threshold fluence of approximately 300~mJ/cm$^2$. Below this threshold, as seen in Fig.~\ref{fig:rouge_bleue_results}(b) at 240~mJ/cm$^2$, no distinct Brillouin frequency can be detected. We also performed control experiments using the glass substrate alone, which is transparent at both the one-photon 780~nm and two-photon 390~nm absorption wavelengths. These tests confirmed that the substrate itself does not contribute to the Brillouin signal. Thus, we attribute the observed nonlinear optical absorption solely to the thin film, reinforcing the conclusion that the Brillouin signals are a result of the nonlinear 2PA properties of the organoruthenium compound, in line with our expectations \cite{Triadon2018}. 

\begin{figure}[!htb]
    \centering
    \includegraphics[width=0.478\textwidth]{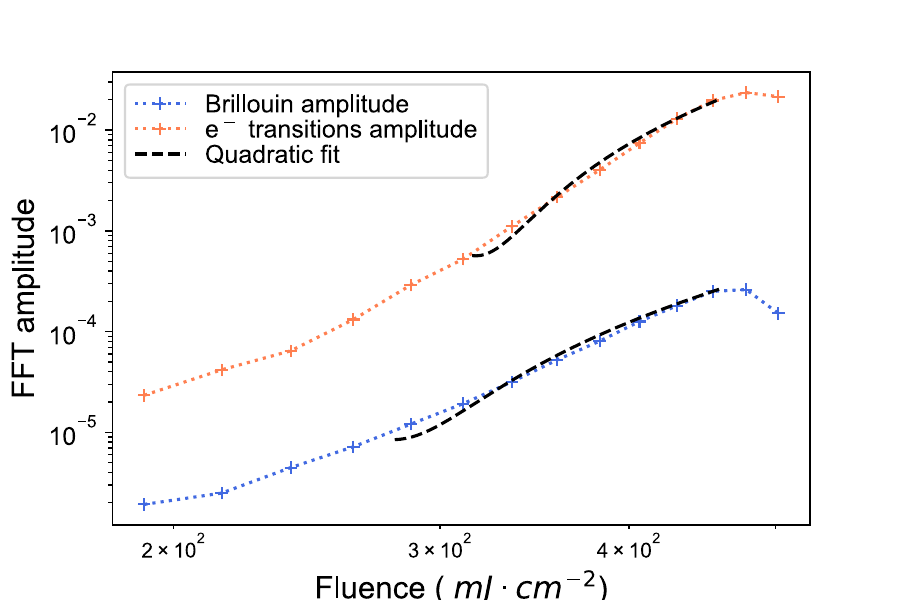}
    \caption{Brillouin and electronic transitions background amplitudes of Fig.~(\ref{fig:rouge_bleue_results}), plotted on a logarithmic scale.}
    \label{fig:rouge_bleue_log_log}
\end{figure} 

{\color{black}It is not only the amplitude of the Brillouin oscillations that carries information about the laser excitation process, but also the phase\cite{Vaudel2014,Parpiiev2017}. This marks a significant departure from conventional Brillouin spectroscopy, where the phase is typically random and lacks informative value. In the time-domain Brillouin scattering experiments presented here, the Brillouin phase is directly linked to the characteristic timescale of the onset of laser-induced acoustic excitation. As a result, Brillouin oscillations provide two complementary insights: the amplitude reflects changes in the efficiency of the excitation process, while the phase reveals shifts in the temporal dynamics of energy deposition. The observed rather constant Brillouin phase in Fig.~\ref{fig:rouge_bleue_results}(d), indicates that the excitation mechanism remains fundamentally stable across the explored fluence range, up to the threshold where optical damage becomes apparent.}

These results conclusively demonstrate that the investigated organoruthenium compound is effectively photothermally and photoacoustically activated through 2PA within the NIR-I transparency window, validating its potential use for nonlinear PA or PT applications.

\begin{figure*}[!htb]
\centering
    \includegraphics[width=\textwidth]{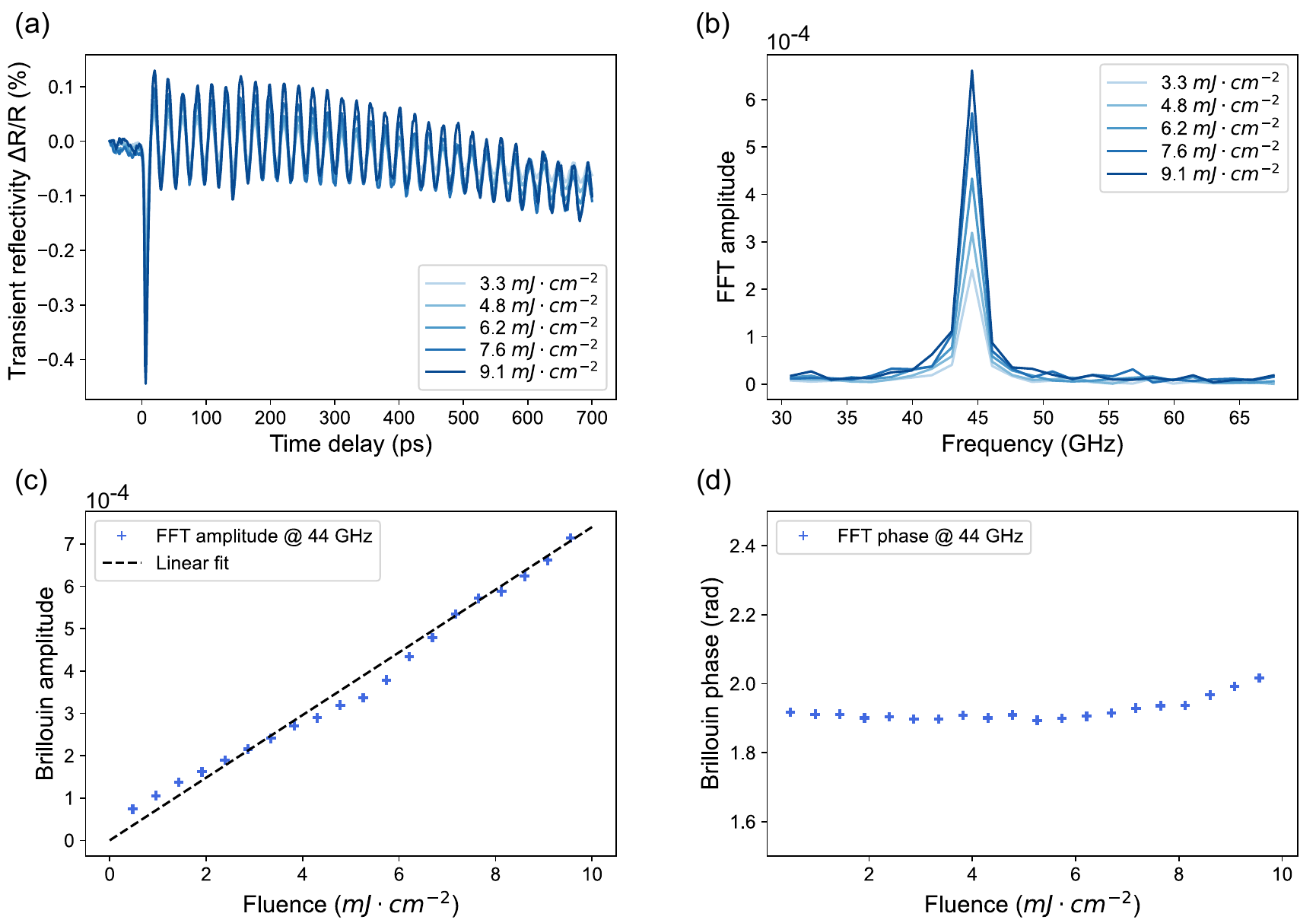}
    \caption{TDBS data at varying pump fluences at a pump wavelength of 390~nm. (a) Time-domain transient reflectivity data. (b) FFT of the the transient reflectivity data after slow background removal. (c) FFT amplitude of the Brillouin peak at 44~GHz versus pump fluence. (d) FFT phase of the Brillouin peak at 44~GHz versus pump fluence.}
    \label{fig:bleue_bleue_results}
\end{figure*}

An essential aspect of TDBS is its sensitivity to acoustic phonon wavepackets on the picosecond timescale. TDBS is not suitable to detect phonons with durations longer than a portion of the Brillouin period \cite{Vaudel2014}. Therefore, the detection of Brillouin oscillations in the glass substrate indicates that the relaxation from the excited state (S$_1$) to the triplet state (T$_1$) is faster than a few picoseconds. If the relaxation would be slower, no Brillouin signal would be observed. This selective sensitivity of TDBS allows it to probe fast relaxation channels while excluding slower ones, which may still manifest in the reflectivity data as a slow electronic transitions background signal, as seen in Fig.~\ref{fig:rouge_bleue_results}(a).

To scrutinize the relationship between the Brillouin signal and the slow electronic transitions background, we plotted Fig.~\ref{fig:rouge_bleue_results}(c) on a logarithmic scale and included the electronic transitions background amplitude, arbitrarily estimated as the average transient reflectivity between 10 and 100~ps. This range avoids the time-zero spike and averages out the Brillouin oscillations, giving a more accurate measure of the thermal baseline. In Fig.~\ref{fig:rouge_bleue_log_log}, the quadratic dependence observed in both the electronic transitions background and the Brillouin amplitude with respect to pump fluence indicates that both channels are connected to 2PA process. This confirms that while Brillouin scattering represents the fast response, the slower background attributed to electronic transitions is also driven by the same 2PA mechanism.

From the theoretical point of view, the laser-excited strain linked to 2PA, derived from Zeuschner \textit{et al.} \cite{Zeuschner}, can be expressed in the form,
\begin{equation}
\label{eqn:strain_nlo}
\eta_{33}^{2PA} = \dfrac{\gamma}{\rho \, C_p} \sqrt{\dfrac{\ln{4}}{\pi}} \, \left(\alpha^{(2)}\dfrac{\mathcal{F}_{2PA}^2}{\tau}\right)
\end{equation}
\noindent where $\eta_{33}^{2PA}$ is the unidirectional longitudinal strain propagating along the sample normal, $\mathcal{F}_{_{2PA}}$ the effective laser fluence for 2PA processes, $\rho$ the film density, $C_p$ the specific heat, $\gamma$ the linear thermal expansion coefficient, $\tau$ the FWHM laser pulse duration, and $\alpha^{(2)}$ the 2PA coefficient. Note that in Eq.~\eqref{eqn:strain_nlo}, the effective fluence $\mathcal{F}_{2PA}$ is linked to the laser fluence in air $\mathcal{F}_0$ through $\mathcal{F}_{2PA} \equiv \mathcal{F}_0 (1-\mathcal{R})$ with $\mathcal{R}$ being the air/film optical reflectivity at the pump wavelength. The strain given by Eq.~\eqref{eqn:strain_nlo} is directly proportional to the amplitude of the Brillouin signal recorded in the glass substrate. To go beyond a qualitative analysis and extract the coefficient $\alpha^{(2)}$ quantitatively from Eq.~\eqref{eqn:strain_nlo}, several key parameters, such as density, heat capacity and thermal expansion, are required. Typically, these quantities are difficult to measure for thin films and can require separate, time-consuming experiments. However, we demonstrate that by performing measurements at two different pump wavelengths, one can bypass the need for these additional measurements.

With this goal, additional experiments have been conducted by changing the pump wavelength from 780~nm (2PA) to 390~nm (1PA), keeping the probe wavelength unchanged at 390~nm. The 390~nm pump is obtained by second harmonic generation in a $\beta$-BBO crystal from the 780~nm beam, for which the efficiency is in the percents range; the pump fluence in this case is a few percents of the fluence reached for 2PA. Although the 1PA pump could not reach the same fluence as the 2PA, Fig.~\ref{fig:bleue_bleue_results}(a) evidences that Brillouin scattering can be detected from a 1PA process. In contrast to Fig.~\ref{fig:rouge_bleue_results}(a), Fig.~\ref{fig:bleue_bleue_results}(a) does not show any pronounced thermal or electronic transitions baseline, and the different TDBS acquisitions are superimposed. The Brillouin frequency in glass is very apparent in the FFT of the time-resolved data, see  Fig.~\ref{fig:bleue_bleue_results}(b). As expected, the evolution of the Brillouin amplitude against pump fluence is linear, see Fig.~\ref{fig:bleue_bleue_results}(c). Moreover, the Brillouin phase in Fig.~\ref{fig:bleue_bleue_results}(d), is constant within the small fluctuations, indicating no modification of the 1PA excitation at this fluence range. 

Notably, the Brillouin amplitude from 1PA in Fig.~\ref{fig:bleue_bleue_results}(c) reaches approximately 1$\times$10$^{21}$ at a fluence of 6~mJ$\cdot$cm$^{-2}$, while for 2PA in Fig.~\ref{fig:rouge_bleue_results}(c), the amplitude reaches around 1$\times$10$^{21}$ at a fluence as high as 450~mJ$\cdot$cm$^{-2}$. This indicates  that 2PA is about 75 times less efficient than 1PA in this context. {\color{black}While this statement might appear as a limitation for using 2PA excitation for performing PA imaging, we must stress here that an even worse efficiency between 1PA and 2PA was stated for fluorescence imaging for which 2PA excitation is clearly recognized as being more advantageous \cite{Dreano2023}. The same will hold for PA imaging. This is  because the lower excitation efficiency of 2PA is largely compensated by specific advantages of this kind of excitation which are: (i) deeper penetration of light into tissues when excitation takes place in proper NIR windows \cite{Dreano2023}, (ii) better resolution achievable in 2PA due to the nonlinear dependance of the response used for imaging \cite{Urban2014,Nedosekin2014,Langer2013} and (iii) lower photodamages to tissues due to the more localized nature of the excitation during sampling \cite{Dreano2023}.} These findings therefore highlight the potential for our technique to be used in screening a wide range of compounds, optimizing those that might show better performance under 2PA conditions.

The analytical expression of the laser-excited strain linked to 1PA, derived from Zeuschner et. al \cite{Zeuschner}, can be expressed in the form,
\begin{equation}
    \eta_{33}^{1PA}=\dfrac{\gamma}{\rho \, C_p \, d}(1-e^{-\alpha^{(1)} d}) \mathcal{F}_{1PA}
    \label{equa:strain_1}
\end{equation}
where $d$ is the film thickness, $\alpha^{(1)}$ the 1PA optical absorption coefficient, and $\mathcal{F}_{_{1PA}}$ the effective fluence of the 1PA pump. Note that Eq.~(\ref{equa:strain_1}) expresses the strain amplitude averaged over the film, while  Eq.~(\ref{eqn:strain_nlo}) expresses the homogeneous -- constant-- strain amplitude in the film. Since the probe wavelength is kept the same for both 1PA and 2PA experiments, the Brillouin amplitudes and the strains in Eq.~(\ref{eqn:strain_nlo}) and Eq.~(\ref{equa:strain_1}) scale linearly for both situations. This linear relationship can be expressed as,
\begin{equation}
\begin{split}
    A_B^{1PA} & = K_{1PA}\, \eta_{33}^{1PA}, \\[0.1cm] 
    A_B^{2PA} & = K_{2PA}\, \eta_{33}^{2PA}.
\end{split}
    \label{coeffAB}
\end{equation}

The Brillouin amplitude $A_B$ in Eq.(\ref{coeffAB}) is related to various physical parameters, encapsulated in the scaling factors $K_{1/2PA}$, which includes the photoelastic coefficient of the glass substrate, the refractive index, the acoustic properties of the film and on the shape of the strain profile excited \cite{Lin1991}. If we make the reasonable assumption that laser excitation in the 1PA case is rather homogeneous through the film thickness, as it is in the 2PA case, then the strain profile generated in both cases should be identical, such that,
\begin{equation}
    K_{1PA} = K_{2PA} = K,
    \label{K}
\end{equation}
This assumption becomes more accurate as the film thickness decreases, simplifying the problem. Under this assumption, there is no need even to run TDBS simulations based on Finite Element Modeling (FEM) used in our previous studies \cite{Chaban2022} to extract $K_{1PA}$ and $K_{2PA}$. This allows for the direct extraction of the nonlinear absorption coefficient $\alpha^{(2)}$ without resorting to complex side-calculations for determining the strain profiles.

Numerically, the quadratic fit of Fig.~\ref{fig:rouge_bleue_results}(c) and of the linear fit of Fig.~\ref{fig:bleue_bleue_results}(c), lead to 
\begin{equation}
\begin{split}
    A_B^{2PA} & = a \, \mathcal{F}_{_{2PA}}^2, \\[0.1cm]
    A_B^{1PA} & = b \, \mathcal{F}_{_{1PA}},
\end{split}
    \label{coeffAB2}    
\end{equation}
%
with $a$\,=\,7.9$\times$10$^{-9}$ mJ$^{-2}\cdot$cm$^4$ and \\ 
$b$\,=\,7.4$\times$10$^{-5}$ mJ$^{-1}\cdot$ cm$^2$. By dividing these coefficients $a$ and $b$ by each other, we obtain from Eq.~(\ref{equa:strain_1}) and Eq.~(\ref{eqn:strain_nlo}), an analytical expression that takes the form,
\begin{equation}
\label{extract}
\dfrac{a}{b} = \sqrt{\dfrac{\ln{4}}{\pi}}\left(\dfrac{\alpha^{(2)}}{\tau}\right)\dfrac{d}{(1-e^{-\alpha^{(1)} d})},
\end{equation}
that no longer contains unknown constants except for the 1PA absorption coefficient $\alpha^{(1)}$, which can be easily measured through widely available techniques. We used the ellipsometry method to extract a value of $\alpha^{(1)}$\,=\,3.72$\times$10$^{4}$ cm$^{-1}$, see Supplementary information. Numerically, we can then calculate the nonlinear optical constant. We obtain: 
\begin{equation}
\label{alpha2}
\alpha^{(2)} = 1.7 \,\, \text{cm/GW}.
\end{equation}
The values obtained for $\alpha^{(2)}$ in our experiments are consistent with results previously reported\cite{Zeuschner,Chaban2022}, supporting the reliability of our approach.
From this experimental value, we calculate using Eq.~(\ref{sigma}) the 2PA cross section expressed in GM at 800~nm wavelength. Considering that the molecular density $N_0$ is of 5.04~$\times$10$^{20}$~cm$^{-3}$ (X-ray value for \textbf{1} in \cite{Triadon2018}) and that 1~GM is equal to 10$^{-50}$~cm$^4\cdot$s, we obtain,
\begin{equation}
\label{sigma2}
\sigma_2 = 81 \,\, \text{GM}.
\end{equation}
Interestingly, this value is approximately one order of magnitude lower than that derived from the conventional z-scan technique. While both femtosecond methods are subject to measurement uncertainties, these uncertainties are well within acceptable limits and below one order of magnitude. This discrepancy highlights the specificity of the time-domain Brillouin approach, which is sensitive to thermoelastic excitation of acoustic waves on a picosecond timescale. Consequently, the coefficient extracted from Brillouin measurements reflects only a portion of the total 2PA coefficient obtained through z-scan, as the latter is not tied to relaxation pathways but rather to the onset of 2PA. Accordingly, we tentatively propose that only approximately one-tenth (81~GM/1075~GM) of the energy needed to excite complex \textbf{1} in its S$_1$ state does actually contribute to heat and acoustic wave generation within the picosecond timescale. This distinction between z-scan and Brillouin emphasizes the unique capabilities of time-domain Brillouin scattering in isolating the fastest dynamics underlying PT/PA processes. Extending this method to a broader range of materials will provide deeper insights into the nonlinear optical properties of potential contrast agents, thereby aiding the development of advanced tools for biomedical imaging.  



\section*{Summary}

We highlight the potential of a slightly modified conventional time-domain Brillouin scattering technique for the all-optical determination of the effective two-photon absorption (2PA) coefficient associated with photothermal  and photoacoustic mechanisms. In this article, we present a streamlined approach based on multi-spectral pump-probe measurements, enabling rapid screening and optimization of the nonlinear absorption properties of nanoscale samples. Using this method, we successfully demonstrated 2PA in a potential contrast agent, an organometallic molecular compound, operating within the biological transparency window NIR-I. These findings pave the way for future research into the nonlinear optical response of organic materials at picosecond timescales, using optical radiation spanning from NIR-I to the NUV.

\section*{Acknowledgments}
The authors would like to thank Ievgeniia Chaban for her valuable scientific advice throughout the study. We also extend our deepest gratitude to Keith Nelson from the Massachusetts Institute of Technology for his generous donation of a complete Coherent Rega system, without which this research would not have been possible. This research was funded by the Agence Nationale de la Recherche under grant ANR-22-CE42-0001 GigaSpin, ANR-22-CE50-0018 PulseCoMeth, as well by Région Bretagne under the SAD grant CHOCONDE and the ARED grant BIFOCROM. We are also grateful for the Australian Research Council (half co-tutelle stipend to A.T.), alongside the financial support provided by the CNRS IRP MAITAI and Rennes Métropole.

\bibliography{references}

\begin{thebibliography}{24}%
\makeatletter
\providecommand \@ifxundefined [1]{%
 \@ifx{#1\undefined}
}%
\providecommand \@ifnum [1]{%
 \ifnum #1\expandafter \@firstoftwo
 \else \expandafter \@secondoftwo
 \fi
}%
\providecommand \@ifx [1]{%
 \ifx #1\expandafter \@firstoftwo
 \else \expandafter \@secondoftwo
 \fi
}%
\providecommand \natexlab [1]{#1}%
\providecommand \enquote  [1]{``#1''}%
\providecommand \bibnamefont  [1]{#1}%
\providecommand \bibfnamefont [1]{#1}%
\providecommand \citenamefont [1]{#1}%
\providecommand \href@noop [0]{\@secondoftwo}%
\providecommand \href [0]{\begingroup \@sanitize@url \@href}%
\providecommand \@href[1]{\@@startlink{#1}\@@href}%
\providecommand \@@href[1]{\endgroup#1\@@endlink}%
\providecommand \@sanitize@url [0]{\catcode `\\12\catcode `\$12\catcode
  `\&12\catcode `\#12\catcode `\^12\catcode `\_12\catcode `\%12\relax}%
\providecommand \@@startlink[1]{}%
\providecommand \@@endlink[0]{}%
\providecommand \url  [0]{\begingroup\@sanitize@url \@url }%
\providecommand \@url [1]{\endgroup\@href {#1}{\urlprefix }}%
\providecommand \urlprefix  [0]{URL }%
\providecommand \Eprint [0]{\href }%
\providecommand \doibase [0]{http://dx.doi.org/}%
\providecommand \selectlanguage [0]{\@gobble}%
\providecommand \bibinfo  [0]{\@secondoftwo}%
\providecommand \bibfield  [0]{\@secondoftwo}%
\providecommand \translation [1]{[#1]}%
\providecommand \BibitemOpen [0]{}%
\providecommand \bibitemStop [0]{}%
\providecommand \bibitemNoStop [0]{.\EOS\space}%
\providecommand \EOS [0]{\spacefactor3000\relax}%
\providecommand \BibitemShut  [1]{\csname bibitem#1\endcsname}%
\let\auto@bib@innerbib\@empty
\bibitem [{\citenamefont {Merkes}\ \emph {et~al.}(2020)\citenamefont {Merkes},
  \citenamefont {Zhu}, \citenamefont {Bahukhandi}, \citenamefont {Rueping},
  \citenamefont {Kiessling},\ and\ \citenamefont {Banala}}]{Merkes2020}%
  \BibitemOpen
  \bibfield  {author} {\bibinfo {author} {\bibfnamefont {J.-M.}\ \bibnamefont
  {Merkes}}, \bibinfo {author} {\bibfnamefont {L.}~\bibnamefont {Zhu}},
  \bibinfo {author} {\bibfnamefont {S.~B.}\ \bibnamefont {Bahukhandi}},
  \bibinfo {author} {\bibfnamefont {M.}~\bibnamefont {Rueping}}, \bibinfo
  {author} {\bibfnamefont {F.}~\bibnamefont {Kiessling}}, \ and\ \bibinfo
  {author} {\bibfnamefont {S.}~\bibnamefont {Banala}},\ }\href@noop {}
  {\bibfield  {journal} {\bibinfo  {journal} {International Journal of
  Molecular Sciences}\ }\textbf {\bibinfo {volume} {21}},\ \bibinfo {pages}
  {3082} (\bibinfo {year} {2020})}\BibitemShut {NoStop}%
\bibitem [{\citenamefont {Hu}\ \emph {et~al.}(2021)\citenamefont {Hu},
  \citenamefont {Prasad},\ and\ \citenamefont {Huang}}]{Hu2021}%
  \BibitemOpen
  \bibfield  {author} {\bibinfo {author} {\bibfnamefont {W.}~\bibnamefont
  {Hu}}, \bibinfo {author} {\bibfnamefont {P.~N.}\ \bibnamefont {Prasad}}, \
  and\ \bibinfo {author} {\bibfnamefont {W.}~\bibnamefont {Huang}},\
  }\href@noop {} {\bibfield  {journal} {\bibinfo  {journal} {Accounts of
  Chemical Research}\ }\textbf {\bibinfo {volume} {54}},\ \bibinfo {pages}
  {697} (\bibinfo {year} {2021})}\BibitemShut {NoStop}%
\bibitem [{\citenamefont {Xu}\ \emph {et~al.}(2022)\citenamefont {Xu},
  \citenamefont {Ye}, \citenamefont {Shen}, \citenamefont {Lam}, \citenamefont
  {Zhao},\ and\ \citenamefont {Zhong~Tang}}]{Xu2022}%
  \BibitemOpen
  \bibfield  {author} {\bibinfo {author} {\bibfnamefont {C.}~\bibnamefont
  {Xu}}, \bibinfo {author} {\bibfnamefont {R.}~\bibnamefont {Ye}}, \bibinfo
  {author} {\bibfnamefont {H.}~\bibnamefont {Shen}}, \bibinfo {author}
  {\bibfnamefont {J.~W.~Y.}\ \bibnamefont {Lam}}, \bibinfo {author}
  {\bibfnamefont {Z.}~\bibnamefont {Zhao}}, \ and\ \bibinfo {author}
  {\bibfnamefont {B.}~\bibnamefont {Zhong~Tang}},\ }\href@noop {} {\bibfield
  {journal} {\bibinfo  {journal} {Angew. Chem. Int. Ed.}\ }\textbf {\bibinfo
  {volume} {61}},\ \bibinfo {pages} {e202204604} (\bibinfo {year}
  {2022})}\BibitemShut {NoStop}%
\bibitem [{\citenamefont {Fu}\ \emph {et~al.}(2019)\citenamefont {Fu},
  \citenamefont {Zhu}, \citenamefont {Song}, \citenamefont {Yang},\ and\
  \citenamefont {Chen}}]{Fu2019}%
  \BibitemOpen
  \bibfield  {author} {\bibinfo {author} {\bibfnamefont {Q.}~\bibnamefont
  {Fu}}, \bibinfo {author} {\bibfnamefont {R.}~\bibnamefont {Zhu}}, \bibinfo
  {author} {\bibfnamefont {J.}~\bibnamefont {Song}}, \bibinfo {author}
  {\bibfnamefont {H.}~\bibnamefont {Yang}}, \ and\ \bibinfo {author}
  {\bibfnamefont {X.}~\bibnamefont {Chen}},\ }\href@noop {} {\bibfield
  {journal} {\bibinfo  {journal} {Advanced Materials}\ }\textbf {\bibinfo
  {volume} {31}},\ \bibinfo {pages} {1805875} (\bibinfo {year}
  {2019})}\BibitemShut {NoStop}%
\bibitem [{\citenamefont {Dréano}\ \emph {et~al.}(2023)\citenamefont
  {Dréano}, \citenamefont {Mongin}, \citenamefont {Paul},\ and\ \citenamefont
  {Humphrey}}]{Dreano2023}%
  \BibitemOpen
  \bibfield  {author} {\bibinfo {author} {\bibfnamefont {M.}~\bibnamefont
  {Dréano}}, \bibinfo {author} {\bibfnamefont {O.}~\bibnamefont {Mongin}},
  \bibinfo {author} {\bibfnamefont {F.}~\bibnamefont {Paul}}, \ and\ \bibinfo
  {author} {\bibfnamefont {M.~G.}\ \bibnamefont {Humphrey}},\ }\href@noop {}
  {\bibfield  {journal} {\bibinfo  {journal} {Australian Journal of Chemistry}\
  }\textbf {\bibinfo {volume} {76}},\ \bibinfo {pages} {130} (\bibinfo {year}
  {2023})}\BibitemShut {NoStop}%
\bibitem [{\citenamefont {Hu}\ \emph {et~al.}(2020)\citenamefont {Hu},
  \citenamefont {Zhang}, \citenamefont {He}, \citenamefont {Baev},
  \citenamefont {Xia}, \citenamefont {Huang},\ and\ \citenamefont
  {Prasad}}]{Hu2020}%
  \BibitemOpen
  \bibfield  {author} {\bibinfo {author} {\bibfnamefont {W.}~\bibnamefont
  {Hu}}, \bibinfo {author} {\bibfnamefont {H.}~\bibnamefont {Zhang}}, \bibinfo
  {author} {\bibfnamefont {G.~S.}\ \bibnamefont {He}}, \bibinfo {author}
  {\bibfnamefont {A.}~\bibnamefont {Baev}}, \bibinfo {author} {\bibfnamefont
  {J.}~\bibnamefont {Xia}}, \bibinfo {author} {\bibfnamefont {W.}~\bibnamefont
  {Huang}}, \ and\ \bibinfo {author} {\bibfnamefont {P.~N.}\ \bibnamefont
  {Prasad}},\ }\href@noop {} {\bibfield  {journal} {\bibinfo  {journal} {ACS
  Photonics}\ }\textbf {\bibinfo {volume} {7}},\ \bibinfo {pages} {3161}
  (\bibinfo {year} {2020})}\BibitemShut {NoStop}%
\bibitem [{\citenamefont {Urban}\ \emph {et~al.}(2014)\citenamefont {Urban},
  \citenamefont {Yi}, \citenamefont {Yakovlev},\ and\ \citenamefont
  {Zhang}}]{Urban2014}%
  \BibitemOpen
  \bibfield  {author} {\bibinfo {author} {\bibfnamefont {B.~E.}\ \bibnamefont
  {Urban}}, \bibinfo {author} {\bibfnamefont {J.}~\bibnamefont {Yi}}, \bibinfo
  {author} {\bibfnamefont {V.}~\bibnamefont {Yakovlev}}, \ and\ \bibinfo
  {author} {\bibfnamefont {H.~F.}\ \bibnamefont {Zhang}},\ }\href@noop {}
  {\bibfield  {journal} {\bibinfo  {journal} {Journal of Biomedical Optics}\
  }\textbf {\bibinfo {volume} {19}},\ \bibinfo {pages} {085001} (\bibinfo
  {year} {2014})}\BibitemShut {NoStop}%
\bibitem [{\citenamefont {Nedosekin}\ \emph {et~al.}(2014)\citenamefont
  {Nedosekin}, \citenamefont {Galanzha}, \citenamefont {Dervishi},
  \citenamefont {Biris},\ and\ \citenamefont {Zharov}}]{Nedosekin2014}%
  \BibitemOpen
  \bibfield  {author} {\bibinfo {author} {\bibfnamefont {D.~A.}\ \bibnamefont
  {Nedosekin}}, \bibinfo {author} {\bibfnamefont {E.~I.}\ \bibnamefont
  {Galanzha}}, \bibinfo {author} {\bibfnamefont {E.}~\bibnamefont {Dervishi}},
  \bibinfo {author} {\bibfnamefont {A.~S.}\ \bibnamefont {Biris}}, \ and\
  \bibinfo {author} {\bibfnamefont {V.~P.}\ \bibnamefont {Zharov}},\
  }\href@noop {} {\bibfield  {journal} {\bibinfo  {journal} {Small}\ }\textbf
  {\bibinfo {volume} {10}},\ \bibinfo {pages} {135} (\bibinfo {year}
  {2014})}\BibitemShut {NoStop}%
\bibitem [{\citenamefont {Langer}\ \emph {et~al.}(2013)\citenamefont {Langer},
  \citenamefont {Bouchal}, \citenamefont {Gr\"{u}n}, \citenamefont
  {Burgholzer},\ and\ \citenamefont {Berer}}]{Langer2013}%
  \BibitemOpen
  \bibfield  {author} {\bibinfo {author} {\bibfnamefont {G.}~\bibnamefont
  {Langer}}, \bibinfo {author} {\bibfnamefont {K.-D.}\ \bibnamefont {Bouchal}},
  \bibinfo {author} {\bibfnamefont {H.}~\bibnamefont {Gr\"{u}n}}, \bibinfo
  {author} {\bibfnamefont {P.}~\bibnamefont {Burgholzer}}, \ and\ \bibinfo
  {author} {\bibfnamefont {T.}~\bibnamefont {Berer}},\ }\href@noop {}
  {\bibfield  {journal} {\bibinfo  {journal} {Optics Express}\ }\textbf
  {\bibinfo {volume} {21}},\ \bibinfo {pages} {22410} (\bibinfo {year}
  {2013})}\BibitemShut {NoStop}%
\bibitem [{\citenamefont {Gu}\ and\ \citenamefont {Zhong}(2023)}]{Gu2013}%
  \BibitemOpen
  \bibfield  {author} {\bibinfo {author} {\bibfnamefont {K.}~\bibnamefont
  {Gu}}\ and\ \bibinfo {author} {\bibfnamefont {H.}~\bibnamefont {Zhong}},\
  }\href@noop {} {\bibfield  {journal} {\bibinfo  {journal} {Light Science \&
  Applications}\ ,\ \bibinfo {pages} {120}} (\bibinfo {year}
  {2023})}\BibitemShut {NoStop}%
\bibitem [{\citenamefont {Lucas}\ \emph {et~al.}(2022)\citenamefont {Lucas},
  \citenamefont {Sarkar}, \citenamefont {Atlas}, \citenamefont {Linger},
  \citenamefont {Renault}, \citenamefont {Gazeau},\ and\ \citenamefont
  {Gateau}}]{Lucas2022}%
  \BibitemOpen
  \bibfield  {author} {\bibinfo {author} {\bibfnamefont {T.}~\bibnamefont
  {Lucas}}, \bibinfo {author} {\bibfnamefont {M.}~\bibnamefont {Sarkar}},
  \bibinfo {author} {\bibfnamefont {Y.}~\bibnamefont {Atlas}}, \bibinfo
  {author} {\bibfnamefont {C.}~\bibnamefont {Linger}}, \bibinfo {author}
  {\bibfnamefont {G.}~\bibnamefont {Renault}}, \bibinfo {author} {\bibfnamefont
  {F.}~\bibnamefont {Gazeau}}, \ and\ \bibinfo {author} {\bibfnamefont
  {J.}~\bibnamefont {Gateau}},\ }\href@noop {} {\bibfield  {journal} {\bibinfo
  {journal} {Sensors}\ }\textbf {\bibinfo {volume} {22}},\ \bibinfo {pages}
  {6543} (\bibinfo {year} {2022})}\BibitemShut {NoStop}%
\bibitem [{\citenamefont {Wang}\ \emph {et~al.}(2014)\citenamefont {Wang},
  \citenamefont {Li}, \citenamefont {Ding},\ and\ \citenamefont
  {Sun}}]{Wang2014}%
  \BibitemOpen
  \bibfield  {author} {\bibinfo {author} {\bibfnamefont {X.}~\bibnamefont
  {Wang}}, \bibinfo {author} {\bibfnamefont {G.}~\bibnamefont {Li}}, \bibinfo
  {author} {\bibfnamefont {Y.}~\bibnamefont {Ding}}, \ and\ \bibinfo {author}
  {\bibfnamefont {S.}~\bibnamefont {Sun}},\ }\href@noop {} {\bibfield
  {journal} {\bibinfo  {journal} {RSC Advances}\ }\textbf {\bibinfo {volume}
  {4}},\ \bibinfo {pages} {30375} (\bibinfo {year} {2014})}\BibitemShut
  {NoStop}%
\bibitem [{\citenamefont {Roper}\ \emph {et~al.}(2007)\citenamefont {Roper},
  \citenamefont {Ahn},\ and\ \citenamefont {Hoepfner}}]{Roper2007}%
  \BibitemOpen
  \bibfield  {author} {\bibinfo {author} {\bibfnamefont {D.~K.}\ \bibnamefont
  {Roper}}, \bibinfo {author} {\bibfnamefont {W.}~\bibnamefont {Ahn}}, \ and\
  \bibinfo {author} {\bibfnamefont {M.}~\bibnamefont {Hoepfner}},\ }\href@noop
  {} {\bibfield  {journal} {\bibinfo  {journal} {The Journal of Physical
  Chemistry C}\ }\textbf {\bibinfo {volume} {111}},\ \bibinfo {pages} {3636}
  (\bibinfo {year} {2007})}\BibitemShut {NoStop}%
\bibitem [{\citenamefont {Braslavsky}\ and\ \citenamefont
  {Heibel}(1992)}]{Braslavsky1992}%
  \BibitemOpen
  \bibfield  {author} {\bibinfo {author} {\bibfnamefont {S.~E.}\ \bibnamefont
  {Braslavsky}}\ and\ \bibinfo {author} {\bibfnamefont {G.~E.}\ \bibnamefont
  {Heibel}},\ }\href@noop {} {\bibfield  {journal} {\bibinfo  {journal}
  {Chemical Reviews}\ }\textbf {\bibinfo {volume} {92}},\ \bibinfo {pages}
  {1381} (\bibinfo {year} {1992})}\BibitemShut {NoStop}%
\bibitem [{\citenamefont {Terazima}(2002)}]{Terazima2002}%
  \BibitemOpen
  \bibfield  {author} {\bibinfo {author} {\bibfnamefont {M.}~\bibnamefont
  {Terazima}},\ }\href@noop {} {\bibfield  {journal} {\bibinfo  {journal}
  {Bulletin of the Chemical Society of Japan}\ }\textbf {\bibinfo {volume}
  {74}},\ \bibinfo {pages} {595} (\bibinfo {year} {2002})}\BibitemShut
  {NoStop}%
\bibitem [{\citenamefont {Tam}\ and\ \citenamefont {Patel}(1979)}]{Tam1979}%
  \BibitemOpen
  \bibfield  {author} {\bibinfo {author} {\bibfnamefont {A.}~\bibnamefont
  {Tam}}\ and\ \bibinfo {author} {\bibfnamefont {C.}~\bibnamefont {Patel}},\
  }\href@noop {} {\bibfield  {journal} {\bibinfo  {journal} {Nature}\ }\textbf
  {\bibinfo {volume} {280}},\ \bibinfo {pages} {304} (\bibinfo {year}
  {1979})}\BibitemShut {NoStop}%
\bibitem [{\citenamefont {Chaban}\ \emph {et~al.}(2022)\citenamefont {Chaban},
  \citenamefont {Deska}, \citenamefont {Privault}, \citenamefont {Trzop},
  \citenamefont {Lorenc}, \citenamefont {Kooi}, \citenamefont {Nelson},
  \citenamefont {Samoc}, \citenamefont {Matczyszyn},\ and\ \citenamefont
  {Pezeril}}]{Chaban2022}%
  \BibitemOpen
  \bibfield  {author} {\bibinfo {author} {\bibfnamefont {I.}~\bibnamefont
  {Chaban}}, \bibinfo {author} {\bibfnamefont {R.}~\bibnamefont {Deska}},
  \bibinfo {author} {\bibfnamefont {G.}~\bibnamefont {Privault}}, \bibinfo
  {author} {\bibfnamefont {E.}~\bibnamefont {Trzop}}, \bibinfo {author}
  {\bibfnamefont {M.}~\bibnamefont {Lorenc}}, \bibinfo {author} {\bibfnamefont
  {S.~E.}\ \bibnamefont {Kooi}}, \bibinfo {author} {\bibfnamefont {K.~A.}\
  \bibnamefont {Nelson}}, \bibinfo {author} {\bibfnamefont {M.}~\bibnamefont
  {Samoc}}, \bibinfo {author} {\bibfnamefont {K.}~\bibnamefont {Matczyszyn}}, \
  and\ \bibinfo {author} {\bibfnamefont {T.}~\bibnamefont {Pezeril}},\
  }\href@noop {} {\bibfield  {journal} {\bibinfo  {journal} {Nano Letters}\
  }\textbf {\bibinfo {volume} {22}},\ \bibinfo {pages} {4362} (\bibinfo {year}
  {2022})}\BibitemShut {NoStop}%
\bibitem [{\citenamefont {Triadon}\ \emph {et~al.}(2018)\citenamefont
  {Triadon}, \citenamefont {Grelaud}, \citenamefont {Richy}, \citenamefont
  {Mongin}, \citenamefont {Moxey}, \citenamefont {Dixon}, \citenamefont {Yang},
  \citenamefont {Wang}, \citenamefont {Barlow}, \citenamefont
  {Rault-Berthelot}, \citenamefont {Cifuentes}, \citenamefont {Humphrey},\ and\
  \citenamefont {Paul}}]{Triadon2018}%
  \BibitemOpen
  \bibfield  {author} {\bibinfo {author} {\bibfnamefont {A.}~\bibnamefont
  {Triadon}}, \bibinfo {author} {\bibfnamefont {G.}~\bibnamefont {Grelaud}},
  \bibinfo {author} {\bibfnamefont {N.}~\bibnamefont {Richy}}, \bibinfo
  {author} {\bibfnamefont {O.}~\bibnamefont {Mongin}}, \bibinfo {author}
  {\bibfnamefont {G.~J.}\ \bibnamefont {Moxey}}, \bibinfo {author}
  {\bibfnamefont {I.~M.}\ \bibnamefont {Dixon}}, \bibinfo {author}
  {\bibfnamefont {X.}~\bibnamefont {Yang}}, \bibinfo {author} {\bibfnamefont
  {G.}~\bibnamefont {Wang}}, \bibinfo {author} {\bibfnamefont {A.}~\bibnamefont
  {Barlow}}, \bibinfo {author} {\bibfnamefont {J.}~\bibnamefont
  {Rault-Berthelot}}, \bibinfo {author} {\bibfnamefont {M.~P.}\ \bibnamefont
  {Cifuentes}}, \bibinfo {author} {\bibfnamefont {M.~G.}\ \bibnamefont
  {Humphrey}}, \ and\ \bibinfo {author} {\bibfnamefont {F.}~\bibnamefont
  {Paul}},\ }\href@noop {} {\bibfield  {journal} {\bibinfo  {journal}
  {Organometallics}\ }\textbf {\bibinfo {volume} {37}},\ \bibinfo {pages}
  {2245} (\bibinfo {year} {2018})}\BibitemShut {NoStop}%
\bibitem [{\citenamefont {Liaros}\ and\ \citenamefont
  {Fourkas}(2017)}]{Liaros2017}%
  \BibitemOpen
  \bibfield  {author} {\bibinfo {author} {\bibfnamefont {N.}~\bibnamefont
  {Liaros}}\ and\ \bibinfo {author} {\bibfnamefont {J.~T.}\ \bibnamefont
  {Fourkas}},\ }\href@noop {} {\bibfield  {journal} {\bibinfo  {journal} {Laser
  \& Photonics Reviews}\ }\textbf {\bibinfo {volume} {11}},\ \bibinfo {pages}
  {1700106} (\bibinfo {year} {2017})}\BibitemShut {NoStop}%
\bibitem [{\citenamefont {Brillouin}(1922)}]{Brillouin_scattering}%
  \BibitemOpen
  \bibfield  {author} {\bibinfo {author} {\bibfnamefont {L.}~\bibnamefont
  {Brillouin}},\ }\href@noop {} {\emph {\bibinfo {title}
  {\href{https://doi.org/10.1051/anphys/192209170088}{Diffusion de la lumière
  et des rayons X par un corps transparent homogène, influence de l'agitation
  thermique}}}}\ (\bibinfo  {publisher} {Masson et Cie, Paris},\ \bibinfo
  {year} {1922})\BibitemShut {NoStop}%
\bibitem [{\citenamefont {Vaudel}\ \emph {et~al.}(2014)\citenamefont {Vaudel},
  \citenamefont {Pezeril}, \citenamefont {Lomonosov}, \citenamefont {Lejman},
  \citenamefont {Ruello},\ and\ \citenamefont {Gusev}}]{Vaudel2014}%
  \BibitemOpen
  \bibfield  {author} {\bibinfo {author} {\bibfnamefont {G.}~\bibnamefont
  {Vaudel}}, \bibinfo {author} {\bibfnamefont {T.}~\bibnamefont {Pezeril}},
  \bibinfo {author} {\bibfnamefont {A.}~\bibnamefont {Lomonosov}}, \bibinfo
  {author} {\bibfnamefont {M.}~\bibnamefont {Lejman}}, \bibinfo {author}
  {\bibfnamefont {P.}~\bibnamefont {Ruello}}, \ and\ \bibinfo {author}
  {\bibfnamefont {V.}~\bibnamefont {Gusev}},\ }\href@noop {} {\bibfield
  {journal} {\bibinfo  {journal} {Physical Review B}\ }\textbf {\bibinfo
  {volume} {90}},\ \bibinfo {pages} {014302} (\bibinfo {year}
  {2014})}\BibitemShut {NoStop}%
\bibitem [{\citenamefont {Parpiiev}\ \emph {et~al.}(2017)\citenamefont
  {Parpiiev}, \citenamefont {Servol}, \citenamefont {Lorenc}, \citenamefont
  {Chaban}, \citenamefont {Lefort}, \citenamefont {Collet}, \citenamefont
  {Cailleau}, \citenamefont {Ruello}, \citenamefont {Daro}, \citenamefont
  {Chastanet},\ and\ \citenamefont {Pezeril}}]{Parpiiev2017}%
  \BibitemOpen
  \bibfield  {author} {\bibinfo {author} {\bibfnamefont {T.}~\bibnamefont
  {Parpiiev}}, \bibinfo {author} {\bibfnamefont {M.}~\bibnamefont {Servol}},
  \bibinfo {author} {\bibfnamefont {M.}~\bibnamefont {Lorenc}}, \bibinfo
  {author} {\bibfnamefont {I.}~\bibnamefont {Chaban}}, \bibinfo {author}
  {\bibfnamefont {R.}~\bibnamefont {Lefort}}, \bibinfo {author} {\bibfnamefont
  {E.}~\bibnamefont {Collet}}, \bibinfo {author} {\bibfnamefont
  {H.}~\bibnamefont {Cailleau}}, \bibinfo {author} {\bibfnamefont
  {P.}~\bibnamefont {Ruello}}, \bibinfo {author} {\bibfnamefont
  {N.}~\bibnamefont {Daro}}, \bibinfo {author} {\bibfnamefont {G.}~\bibnamefont
  {Chastanet}}, \ and\ \bibinfo {author} {\bibfnamefont {T.}~\bibnamefont
  {Pezeril}},\ }\href@noop {} {\bibfield  {journal} {\bibinfo  {journal}
  {Applied Physics Letters}\ }\textbf {\bibinfo {volume} {111}},\ \bibinfo
  {pages} {151901} (\bibinfo {year} {2017})}\BibitemShut {NoStop}%
\bibitem [{\citenamefont {Zeuschner}\ \emph {et~al.}(2020)\citenamefont
  {Zeuschner}, \citenamefont {Pudell}, \citenamefont {von Reppert},
  \citenamefont {Deb}, \citenamefont {Popova}, \citenamefont {Keller},
  \citenamefont {Rössle}, \citenamefont {Herzog},\ and\ \citenamefont
  {Bargheer}}]{Zeuschner}%
  \BibitemOpen
  \bibfield  {author} {\bibinfo {author} {\bibfnamefont {S.~P.}\ \bibnamefont
  {Zeuschner}}, \bibinfo {author} {\bibfnamefont {J.-E.}\ \bibnamefont
  {Pudell}}, \bibinfo {author} {\bibfnamefont {A.}~\bibnamefont {von Reppert}},
  \bibinfo {author} {\bibfnamefont {M.}~\bibnamefont {Deb}}, \bibinfo {author}
  {\bibfnamefont {E.}~\bibnamefont {Popova}}, \bibinfo {author} {\bibfnamefont
  {N.}~\bibnamefont {Keller}}, \bibinfo {author} {\bibfnamefont
  {M.}~\bibnamefont {Rössle}}, \bibinfo {author} {\bibfnamefont
  {M.}~\bibnamefont {Herzog}}, \ and\ \bibinfo {author} {\bibfnamefont
  {M.}~\bibnamefont {Bargheer}},\ }\href@noop {} {\bibfield  {journal}
  {\bibinfo  {journal} {Physical Review Research 2}\ }\textbf {\bibinfo
  {volume} {2}},\ \bibinfo {pages} {022013} (\bibinfo {year}
  {2020})}\BibitemShut {NoStop}%
\bibitem [{\citenamefont {Lin}\ \emph {et~al.}(1991)\citenamefont {Lin},
  \citenamefont {Stoner}, \citenamefont {Maris},\ and\ \citenamefont
  {Tauc}}]{Lin1991}%
  \BibitemOpen
  \bibfield  {author} {\bibinfo {author} {\bibfnamefont {H.}~\bibnamefont
  {Lin}}, \bibinfo {author} {\bibfnamefont {R.~J.}\ \bibnamefont {Stoner}},
  \bibinfo {author} {\bibfnamefont {H.~J.}\ \bibnamefont {Maris}}, \ and\
  \bibinfo {author} {\bibfnamefont {J.}~\bibnamefont {Tauc}},\ }\href@noop {}
  {\bibfield  {journal} {\bibinfo  {journal} {Journal of Applied Physics}\
  }\textbf {\bibinfo {volume} {69}},\ \bibinfo {pages} {3816} (\bibinfo {year}
  {1991})}\BibitemShut {NoStop}%
\end{thebibliography}%

\end{document}


\title{Quantification of Ultrafast Nonlinear Photothermal and Photoacoustic Effects in Molecular Thin Films via Time-Domain Brillouin Scattering: \\ Supplemental Information}

\author{Valentin Cherruault}
\affiliation{Institut de Physique de Rennes, UMR CNRS 6251, Université de Rennes, 35042 Rennes, France}
\author{Franck Camerel}
\affiliation{Institut des Sciences Chimiques de Rennes, UMR CNRS 62226, Université de Rennes, 35042 Rennes, France}
\author{Julien Morin}
\affiliation{Institut de Physique de Rennes, UMR CNRS 6251, Université de Rennes, 35042 Rennes, France}
\author{Amédée Triadon}
\affiliation{Institut des Sciences Chimiques de Rennes, UMR CNRS 62226, Université de Rennes, 35042 Rennes, France}
\affiliation{Research School of Chemistry, Australian National University, Canberra, ACT 2601, Australia}
\author{Nicolas Godin}
\affiliation{Institut de Physique de Rennes, UMR CNRS 6251, Université de Rennes, 35042 Rennes, France}
\author{Ronan Lefort}
\affiliation{Institut de Physique de Rennes, UMR CNRS 6251, Université de Rennes, 35042 Rennes, France}
\author{Olivier Mongin}
\affiliation{Institut des Sciences Chimiques de Rennes, UMR CNRS 62226, Université de Rennes, 35042 Rennes, France}
\author{Jean-François Bergamini}
\affiliation{Institut des Sciences Chimiques de Rennes, UMR CNRS 62226, Université de Rennes, 35042 Rennes, France}
\author{Antoine Vacher}
\affiliation{Institut des Sciences Chimiques de Rennes, UMR CNRS 62226, Université de Rennes, 35042 Rennes, France}
\author{Mark G. Humphrey}
\affiliation{Research School of Chemistry, Australian National University, Canberra, ACT 2601, Australia}
\author{Maciej Lorenc}
\email{maciej.lorenc@cnrs.fr}
\affiliation{Institut de Physique de Rennes, UMR CNRS 6251, Université de Rennes, 35042 Rennes, France}
\author{Frederic Paul}
\email{frederic.paul@cnrs.fr}
\affiliation{Institut des Sciences Chimiques de Rennes, UMR CNRS 62226, Université de Rennes, 35042 Rennes, France}
\author{Thomas Pezeril}
\email{thomas.pezeril@cnrs.fr}
\affiliation{Institut de Physique de Rennes, UMR CNRS 6251, Université de Rennes, 35042 Rennes, France}

\date{\today}		

\maketitle

\begin{figure}[!htb]
\centering
\includegraphics[width =11.5cm]{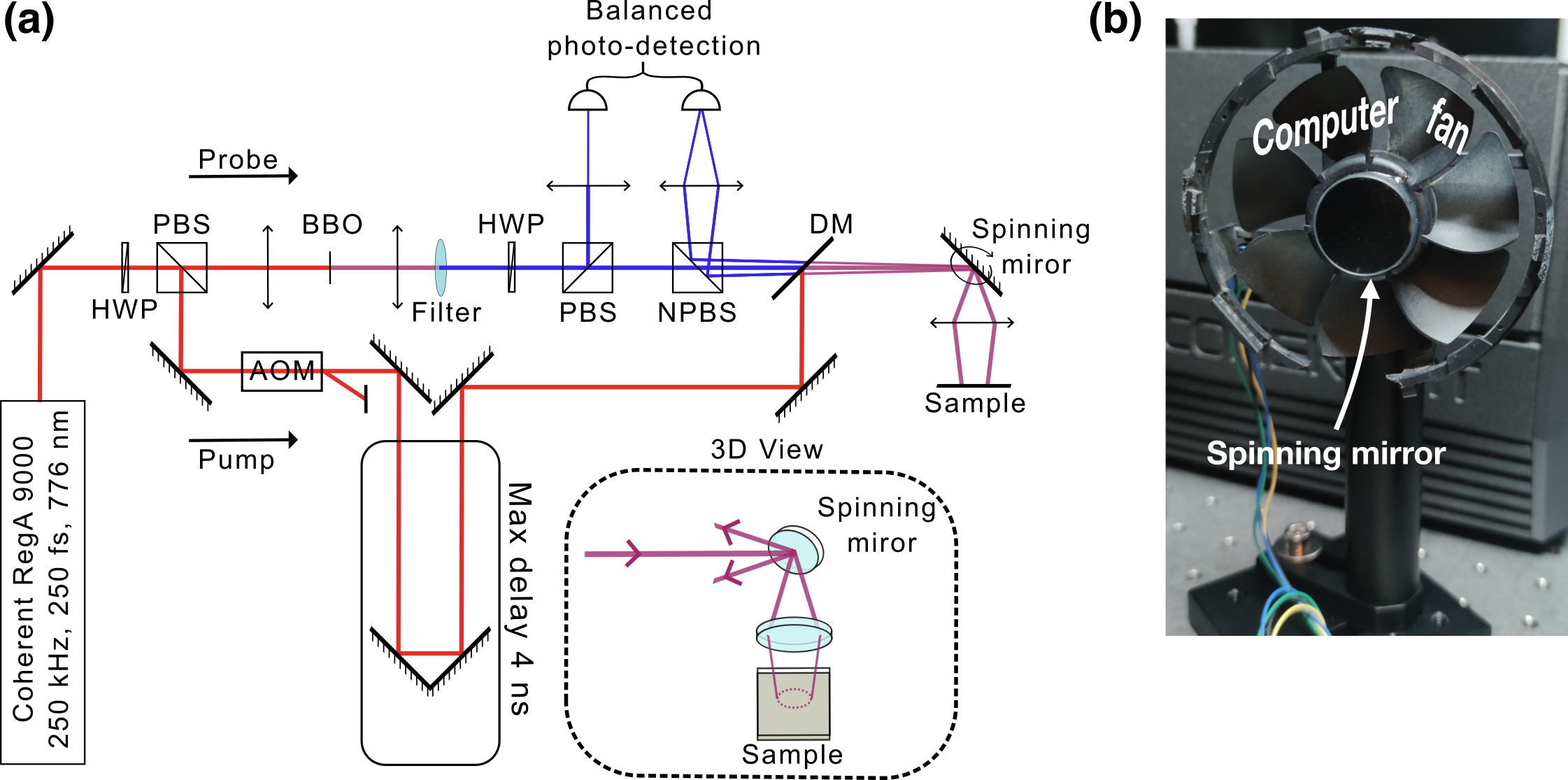}
\caption{(a) Experimental pump/probe setup. HWP: Half Wave Plate, AOM: Acousto-Optic Modulator, BBO: $\beta$-BaB$_2$O$_4$, PBS: Polarizing BeamSplitter, NPBS: Non-Polarizing BeamSplitter, DM: Dichroic mirror. (b) Photo of the spinning wedge mirror unit made of a metallic mirror glued on a computer fan.}
\label{fig:setup}
\end{figure}

\section*{Experimental setup}

The experimental setup is presented in Fig.~\ref{fig:setup}(a). It is an optical pump-probe setup based on a Coherent Rega amplified femtosecond laser. The central wavelength of the laser system is set at 780~nm, the repetition rate at 250~kHz, the pulse duration at 250~fs FWHM and the maximum pulse energy is 4~µJ. The laser is split into a pump beam and a probe beam. The pump passes through an acousto-optic modulator that provides the reference frequency (25~kHz) for the synchronous lock-in detection. The delay line is used to tune the delay between the pump and probe beams in order to time-resolve the acoustic propagation in the glass substrate. The probe is frequency doubled to 380~nm and split into a reference probe and a signal probe for balanced photodetection. The pump and probe beams are recombined using a dichroic mirror (DM) and aligned to be perfectly collinear before being directed onto a wedged spinning mirror unit. This unit reflects the beams and focuses them onto the sample surface through a $\times$10 Mitutoyo microscope objective. The wedged spinning mirror, though low-cost, is a highly effective and practical component. It is constructed from a round metallic mirror affixed to the flat surface of the helices of a standard computer fan, see Fig.~\ref{fig:setup}(b). Powered by a controllable DC voltage source, the fan spins the mirror at approximately 50~Hz.
Due to minor imperfections in the gluing process, the mirror surface is slightly tilted, introducing a wedge angle. As a result, the reflected laser beams describe a conical path whose apex angle corresponds to the wedge-induced deviation. The spinning mirror is positioned close to the microscope objective to ensure the full laser cone passes through without clipping.
Consequently, both the pump and probe beams trace a circular path on the sample surface with a diameter of approximately 200~$\mu$m. The probe beam reflected from the sample, which retains a conical profile, is then redirected to a photodiode for optical detection. To capture the probe beam efficiently, a 10~cm focal length lens is used to re-image the tip of the reflected cone onto the active area of a pin photodiode. This configuration ensures that the beam does not merely sweep across the detector surface, but remains consistently detected, see Fig.~\ref{fig:setup}(a).

In the case of 1PA excitation, the pump is frequency doubled after the delay line and both beams of same wavelengths are re-combined with a D-shaped mirror instead of a dichroic mirror. We did not use the spinning mirror for the identical pump and probe wavelengths configuration. The spinning mirror was crucial for the 2PA excitation at high pump energy, to avoid cumulative heating and optical damage to the sample.

\section*{Sample characterization}
\begin{figure}[!htb]
    \centering
    \includegraphics[width=0.5\linewidth]{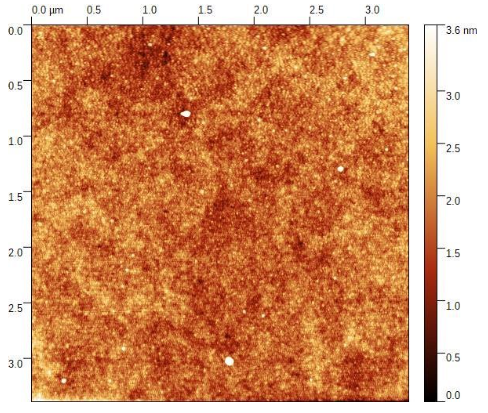}
    \caption{AFM imagery of the organoruthenium thin film spin-coated on a glass SiO$_2$ substrate.}
    \label{fig:AFM}
\end{figure}

We recall that the investigated sample is an organoruthenium complex deposited on a glass substrate through spincoating at 1000~rpm. Several standard sample characterizations were performed. Atomic force microscopy (AFM) imagery was performed and enabled us to check the organoruthenium thin film homogeneous density, as shown in Fig.~\ref{fig:AFM}. The AFM was used as well to precisely measure the film thickness $d$ of $\sim$100 $\pm$ 5~nm.

\begin{figure}[!htb]
    \centering
    \includegraphics[width=\linewidth]{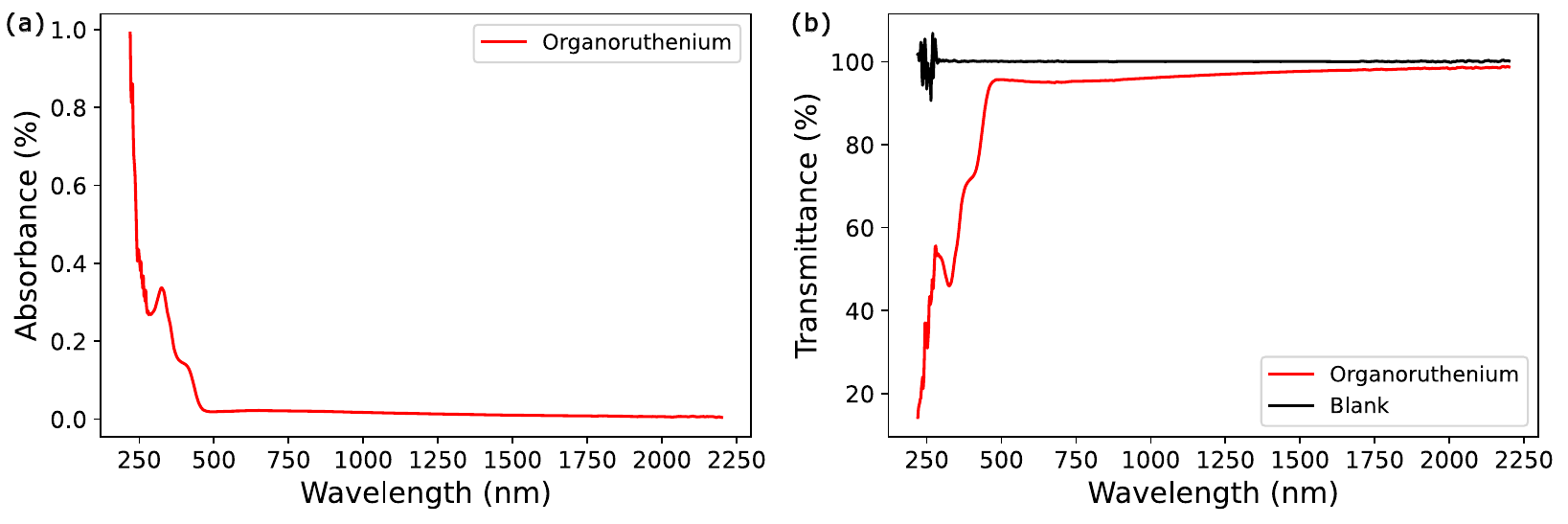}
    \caption{Organoruthenium solid-state UV‐vis-NIR absorption (a) and total transmission spectra (b).}
    \label{fig:absTransSI}
\end{figure}

The determination of the UV‐vis-NIR absorption and total transmission spectra was obtained using a Shimadzu UV‐3600plus spectrophotometer with an integration sphere attachment ISR‐603 using a barium sulfate surface (BaSO$_4$, Nacalai Tesque) as a reflective surface. The reference data were performed on a blank glass substrate. The spectra are shown in Fig~\ref{fig:absTransSI}.

\pagebreak

These measurements confirm the sample transparency of the organoruthenium from 500~nm to mid-IR but, they a not reliable for the extraction of the organoruthenium 1PA optical absorption coefficient. Complementary optical investigations were made with an ellipsometer to determine the real part and imaginary part of the refractive index of the organoruthenium thin film. A set of ellipsometric measurements were made on another batch of organoruthenium thin films deposited on a silicon substrate, which is the most suited reference substrate for ellipsometry. The strong absorption of the silicon substrate, in the investigated range of wavelengths, allows us to consider the substrate as semi-infinite in the modeling. A preliminary ellipsometric measurement on the blank silicon substrate enabled us to determine that the silicon wafer has a SiO$_2$ passivation layer of 7.70~nm.



\begin{figure}[!htb]
    \centering
    \includegraphics[width=0.5\linewidth]{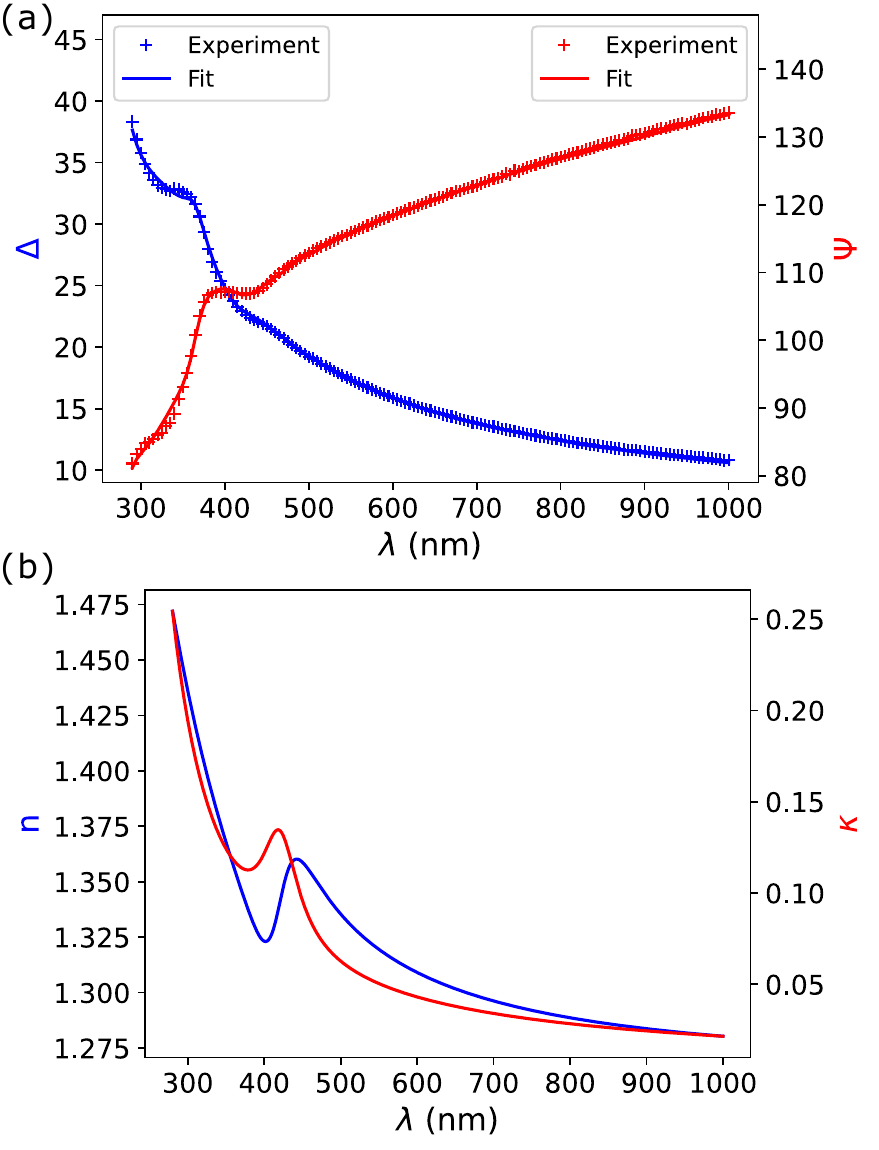}
    \caption{(a) Ellipsometric parameters $\Delta$ and $\Psi$ obtained experimentally and numerical fit. (b) Extraction of the parameters n and $\kappa$ of the organoruthenium film.}
    \label{fig:data_ellipso}
\end{figure}

The ellipsometry data obtained on the organoruthenium thin film are shown in  Fig.~\ref{fig:data_ellipso}. To fit the experimental results, we use the database to model the substrate and the passivation layer, and we used a two-oscillator dispersion model for the organic layer defined as following,
\begin{equation}
    \centering
    \label{Eq:2oscillators}
    \varepsilon(\omega)=\varepsilon_\infty+\sum_{j=1}^{2}\dfrac{f_j \cdot \omega_j^2}{\omega_j^2-\omega^2+i\gamma_j\omega}
\end{equation}
where $\varepsilon$ is the dielectric permittivity, $\varepsilon_\infty$ the high frequency permittivity, $f_j$ the strength of the j$^{th}$ oscillator, $\omega_j$ the resonance frequency of the j$^{th}$ oscillator, $\omega$ the optical pulsation, and $\gamma_j$ the damping factor of the j$^{th}$ oscillator. The numerical calculation of the model parameters provides the values given in Tab.~\ref{tab:2ocillatorsParam}. The choice of the two-oscillator model is due to the two absorption peaks observed in Fig. \ref{fig:absTransSI}. As shown in Fig.~\ref{fig:data_ellipso}(a), the modeling perfectly fits the experimental data. The very good quality of the fit allows us to be confident in the determination of the organoruthenium optical parameters. Finally, the refractive index and the optical absorption of the organoruthenium compound can be extracted through the equation $\Tilde{n}=n+i\kappa = \sqrt{\epsilon(\omega)}$, see Fig. \ref{fig:data_ellipso}(b). Specifically to our needs, we extract the linear absorption coefficient $\alpha^{(1)}$= $4\pi \kappa/\lambda$ = 37235.0~cm$^{-1}$, that is required  to calculate the nonlinear absorption coefficient $\alpha^{(2)}$.

\begin{table}[!htb]
    \centering
    \begin{tabular}{c|c}
    Parameter            & Value \\
    \hline
    $\varepsilon_\infty$ & 0.8868633 \\
    $f_1$                & 0.0213638 \\
    $\omega_1$           & 2.9450393 \\
    $\gamma_1$           & 0.3978620 \\
    $f_2$                & 0.6964802 \\
    $\omega_2$           & 5.7832460 \\
    $\gamma_2$           & 1.8927698
    \end{tabular}
    \caption{Parameters of the two-oscillator model used to match the optical dispersion of the organoruthenium thin film deposited on silicon.}
    \label{tab:2ocillatorsParam}
\end{table}